\documentclass[preprint2]{aastex63}

\usepackage[T1]{fontenc}

\usepackage{bm}        
\usepackage{amssymb, amsmath}   
\usepackage{cancel}
\usepackage{color}

\newcommand\aastex{AAS\TeX}

\DeclareMathAlphabet{\mathsfit}{\encodingdefault}{\sfdefault}{m}{sl}
\SetMathAlphabet{\mathsfit}{bold}{\encodingdefault}{\sfdefault}{bx}{sl}

\newcommand{\vect}[1]{\bm{#1}}

\received{****}
\revised{****}
\accepted{****}
\submitjournal{ApJ}

\shorttitle{\aastex\ Modeling Electron Acceleration in Sep. 10th Event}
\shortauthors{Li et al.}

\begin{document}

\title{Modeling Electron Acceleration and Transport in the Early Impulsive Phase of the 2017 September 10 Solar Flare}

\correspondingauthor{Xiaocan Li}
\email{Xiaocan.Li@dartmouth.edu}

\author[0000-0001-5278-8029]{Xiaocan Li}
\affil{Dartmouth College, Hanover, NH 03750 USA}

\author[0000-0003-4315-3755]{Fan Guo}
\affiliation{Los Alamos National Laboratory, Los Alamos, NM 87545, USA}
\author[0000-0002-0660-3350]{Bin Chen}
\affiliation{Center for Solar-Terrestrial Research, New Jersey Institute of Technology, 3L King Jr. Blvd., Newark, NJ 07102-1982, USA}
\author[0000-0002-9258-4490]{Chengcai Shen}
\affiliation{Harvard-Smithsonian Center for Astrophysics, 60, Garden Street, Cambridge, MA, 02138, USA}
\author[0000-0001-7092-2703]{Lindsay Glesener}
\affiliation{University of Minnesota, Minneapolis, MN, USA}

\begin{abstract}
  The X8.2-class limb flare on September 10, 2017 is among the best studied solar flare events owing to its great similarity to the standard flare model and the broad coverage by multiple spacecraft and ground-based observations. These multiwavelength observations indicate that electron acceleration and transport are efficient in the reconnection and flare looptop regions. However, there lacks a comprehensive model for explaining and interpreting the multi-faceted observations. In this work, we model the electron acceleration and transport in the early impulsive phase of this flare. We solve the Parker transport equation that includes the primary acceleration mechanism during magnetic reconnection in the large-scale flare region modeled by MHD simulations. We find that electrons are accelerated up to several MeV and fill a large volume of the reconnection region, similar to the observations shown in microwaves. The electron spatial distribution and spectral shape in the looptop region agree well with those derived from the microwave and hard X-ray emissions before magnetic islands grow large and dominate the acceleration. Future emission modelings using the electron maps will enable direct comparison with microwave and hard X-ray observations. These results shed new light on the electron acceleration and transport in a broad region of solar flares within a data-constrained realistic flare geometry.
\end{abstract}


\section{Introduction}
Solar flares are among the most remarkable phenomena in the solar system. A large amount of magnetic energy is abruptly released and converted into nonthermal electrons and subsequent emissions~\citep{Lin1976Nonthermal, Emslie2012ApJ,Aschwanden2016ApJ,Warmuth2020}. Understanding electron acceleration and associated nonthermal emissions are of central importance to solar flare research. While magnetic reconnection has been commonly recognized as the driver for magnetic energy release and electron acceleration in solar flares, a complete model that explains the nonthermal electron acceleration and associated emissions remains elusive.

In the standard CSHKP model of eruptive flares~\citep{Carmichael64,Sturrock66,Hirayama74,Kopp76}, oppositely directed magnetic field lines meet at a diffusion region where they break and rejoin, driving energy release and particle acceleration. A fraction of energetic particles escape to the interplanetary space along the open field lines and produce impulsive solar energetic particle (SEP) events~\citep{Reames99}. Some produce emissions at or above the flare arcade, and those precipitating into the dense chromosphere produce emissions at footpoints. This picture has been verified through decades of dedicated observational studies~\citep[e.g.,][]{Masuda1994Loop,Sui2003Evidence,Krucker2008,Krucker2010Measure,Liu2013Plasmoid,Gary2018Microwave,Chen2020Measurement,Chen2020ApJ}. 

One of the most studied events in recent years is the X8.2-class flare on 2017 September 10 (SOL2017-09-10), which was observed at multiple wavelengths and showed excellent agreement with the standard model~\citep[see][and references therein]{Chen2020ApJ}. These observations have shown a large-scale current sheet trailing the erupted flux rope, which plays a key role in driving the successful coronal mass ejection and associated particle acceleration. Both reconnection inflow and bi-directional outflow have been identified in different flare phases~\citep{Cheng2018ApJ,Longcope2018ApJ,Yan2018ApJ,Yu2020ApJ}, and detailed analysis has also suggested that the flare region is turbulent~\citep{Cheng2018ApJ,Warren2018ApJ,French2019}. The coronal plasma in the flare region is heated to $\gtrsim25$ Mk~\citep{Warren2018ApJ,Polito2018ApJ}, and a large number of electrons and ions~\citep{Omodei2018ApJ,Kocharov2020ApJ} are accelerated to nonthermal energies. These energetic electrons produced hard X-ray and microwave emissions, which were observed by RHESSI and the Expanded Owens Valley Solar Array (EOVSA). The microwave emissions revealed that nonthermal electrons are present in a broad region of the flare from the reconnection current sheet to the erupting magnetic flux rope \citep{Chen2020Measurement,Chen2020ApJ}, suggesting efficient acceleration and transport of nonthermal electrons. Using the joint multi-wavelength observations and MHD simulations,~\citet{Chen2020Measurement} managed to get accurate flare geometry, magnetic field strength, and nonthermal electron distributions along the reconnection current sheet. They found that most high-energy electrons ($> 300$ keV) turned out to be at the above-the-looptop region. Joint microwave and HXR image spectroscopy suggests that the nonthermal electron spectrum in the coronal acceleration region fits well with a double power-law function~\citep{Chen2021Energetic}. These features of electron acceleration and transport are yet to be reproduced in numerical modelings.

The primary challenge for modeling electron acceleration and transport in solar flares lies in the large scale-separation in solar flares~\citep{Li2021PoP}. While MHD models are adequate for studying the flare geometry and magnetic evolution~\citep[e.g.,][]{Shen2018,Cheung2018Comprehensive,Chen2020Measurement}, they do not include kinetic particle physics and cannot directly study nonthermal particles. Fully kinetic simulations offer a complete and self-consistent description of plasma kinetic physics and provide critical insights into particle acceleration and transport processes during magnetic reconnection \citep[e.g.,][]{Drake2006Electron,Dahlin2014Mechanisms,Dahlin2017Role,Li2015Nonthermally,Li2017Particle,Li2018Roles,Li2019Formation}. However, they cannot describe large-scale dynamics due to the enormous separation between kinetic scales (ion skin depth $d_i \sim 1$ m) and the scale of the solar flare reconnection region ($\sim 10^4$ km). To better interpret the nonthermal emissions in solar flares, one needs macroscopic energetic-particle models to study particle acceleration and transport within a realistic MHD framework that describes the evolving flare geometry. These models should incorporate particle acceleration and transport processes extracted from kinetic simulations.

Several macroscopic models have been proposed for studying particle acceleration in a large-scale reconnection layer. The models based on guiding-center drift motions have attempted to explain nonthermal particle acceleration associated with magnetic islands in the solar wind~\citep{LeRoux2015Kinetic,LeRoux2016Combining,LeRoux2018Self}. Recently, this model has been coupled with MHD simulations and included feedback from the energetic electrons to the MHD flow~\citep{Drake2019,Arnold2019}. The so-called \texttt{kglobal} model has successfully obtained extended power-law spectra in a large-scale reconnection layer~\citep{Arnold2021PRL}. An alternative approach is based on the conservation of the first and second adiabatic invariants~\citep{Drake2013Power,Zank2014Particle,Zank2015Diffusive}. The resulting energetic particle transport equations have also explained the local enhancements of energetic particles in the solar wind~\citep{Zhao2018Unusual,Zhao2019Particle,Adhikari2019Role}. These equations can evolve anisotropic particle distributions in a ``sea'' of magnetic islands (or flux ropes). When particle distributions are nearly isotropic, the Parker transport equation~\citep{Parker1965Passage}, where the first-order acceleration is due to flow compression, can describe the time evolution of the distribution function. This equation has recently been adopted in solar flare studies to model particle acceleration in a single large-scale reconnection layer~\citep{Li2018Large} and flare termination shocks~\citep{Kong2019,Kong2020ApJ,Kong2022Model}. The results show general agreement with certain observed electron spectra and emission signatures in the coronal region.

In this paper, we solve the Parker transport equation with background fields obtained by MHD simulations for the early impulsive phase of the Sep. 10th, 2017 flare. Our goal is to reproduce the observational signatures of electron acceleration and transport in the large-scale reconnection layer. In Section~\ref{sec:num}, we describe the setup for the MHD simulations and the energetic particle transport modeling. In Section~\ref{sec:results}, we present our simulation results. We show that high-energy electrons (up to a few MeV) can fill a significant portion of the reconnection region, and the resulting spectral shapes and electron maps are consistent with the observations. In Section~\ref{sec:con}, we discuss the model setup, model results, and possible future works.

\section{Numerical Simulations}
\label{sec:num}
We adopted the same 2.5-D MHD simulation setup in~\citet{Chen2020Measurement} which follows the standard eruptive flare scenario, in which reconnection occurs in a vertical current sheet induced by the eruption of a magnetic flux rope. Discussions of the detailed data-constrained model setup can be found in Methods of~\citet{Chen2020Measurement}. The time evolution is similar to the Sep. 10th, 2017 flare~\citep[][see Appendix~\ref{app:flare_evolution} for an illustration]{Gary2018Microwave, Chen2020Measurement}. We normalize the simulations by $L_0=300$ Mm, $B_0=900$ Gauss, and $n_0=8.1\times10^{11}\text{ cm}^{-3}$, which is larger than the typical coronal density due to the difficulty of carrying out a low-$\beta$ MHD simulation with a high Lundquist number. The simulation domain is $x\in[-0.25, 0.25]$ and $y\in[0, 1]$. In the MHD runs, $\beta$ is $\sim0.1$. The resulting characteristic Aflv\'en speed is $V_{A0}=2181.2$ km/s, and the Alfv\'en crossing time $\tau_A=L_0/V_\text{A0}\approx137$ s. We introduce a moderate guide field ($B_z/B_0\approx 0.3$) in the simulations. $B_z$ peaks at the flux rope center and decreases rapidly at a greater distance from the flux rope~\citep{Ye2019,Chen2020Measurement}. The setup is consistent with the observations and MHD simulations~\citep[see][and references therein]{Dahlin2021}, showing that the guide field is localized to the filament channel around the polarity inversion line. The guide field around the reconnection sheet becomes weaker as the simulation evolves, especially near the end of the simulations (the end of the early impulsive phase). We expect magnetic energy release and particle acceleration to become more efficient in the main impulsive phase. Under this guide field, the resulting flare geometry can better reproduce the flare geometry and the magnetic field profile than the simulations with a stronger guide field~\citep{Chen2020Measurement}. We carry out two simulations with a uniform resistivity $\eta=10^{-5}$ and $10^{-6}$, resulting a Lundquist number $S=10^5$ and $10^6$, respectively. We use a high-resolution uniform grid instead of applying adaptive mesh refinement in the current sheet~\citep{Chen2020Measurement}. The grid numbers are $n_x \times n_y = 2048\times4096$ when $S=10^5$ and $8192\times16384$ when $S=10^6$. 

Using the time-dependent magnetic field and velocity field provided by MHD simulations, we solve Parker's transport equation,
\begin{equation}
  \frac{\partial f}{\partial t} + (\vect{V}+\vect{V}_d)\cdot\nabla f
  - \frac{1}{3}\nabla\cdot\vect{V}\frac{\partial f}{\partial\ln p}
  = \nabla\cdot(\vect{\kappa}\nabla f) + Q,
\end{equation}
where $f(\vect{x}, p, t)$ is the particle distribution function; $\vect{\kappa}$ is the spatial diffusion coefficient tensor, $\vect{V}$ is the bulk plasma velocity, and $Q$ is the source. According to kinetic simulations, electrons can be injected from the thermal inflow plasma when streaming along the exhaust boundaries or by the strong ideal/non-ideal electric field near the X-points~\citep{Egedal2013Review,Guo2019Determining,Zhang2021Efficient}. Note that the particle drift $\vect{V}_d$ is out of the simulation plane and is not considered in the 2D simulations. We assume the particle distribution to be nearly isotropic based on earlier results of kinetic simulations~\citep{Li2018Roles,Li2019Formation}, which showed that the anisotropy is weak for high-energy electrons due to pitch-angle scattering by the self-generated turbulence in 3D reconnection~\citep{Daughton2011Role,Guo2015Particle,Dahlin2015Electron,Huang2016Turbulent,Li2019Formation,Zhang2021Efficient}. During the simulation, we continuously inject particles (the source $Q$) in regions with large current densities (calculated from the magnetic field) to mimic the continuous particle fluxes brought into the current sheet by the reconnection inflow and injection out of the thermal pool near the reconnection exhaust and X-point\footnote{$Q$ is constant in time in our simulations. In principle, particle injection should depend on the plasma parameters (e.g., plasma $\beta$ and guide field) and reconnection dynamics (e.g., reconnection rate), and requires more kinetic studies.}. Depending on the local magnetic field direction $\vect{b}$, the spatial diffusion is determined by the diffusion coefficient tensor $\kappa_{ij} = \kappa_\perp\delta_{ij} - (\kappa_\perp-\kappa_\parallel)b_ib_j$, where $\kappa_\parallel$ and $\kappa_\perp$ are the diffusion coefficients along and across the magnetic field lines, respectively. According to the quasi-linear theory~\citep{Jokipii1971Propagation,Giacalone1999Transport}, $\kappa_\parallel\approx1.622v^{4/3}L_c^{2/3}/(\Omega_0^{1/3}\sigma^2)$ when magnetic turbulence is fully developed and has an isotropic power spectrum $\sim k^{-5/3}$, where $v$ is the particle speed, $\Omega_0=eB_0/(m_ec)$ is the particle gyrofrequency, $L_c$ is the turbulence correlation length, and $\sigma^2=\left<\delta B^2\right>/B_0^2$ is the turbulence amplitude. In our simulations, we set $\kappa_\parallel=\kappa_{\parallel 0}(v/v_0)^{4/3}(B_0/B)^{1/3}$, where $v_0$ is thermal speed of the initial 10 keV electrons, $\kappa_{\parallel 0}$ is the corresponding parallel diffusion coefficient, and we assume $L_c\approx140$ km and $\sigma^2=1$. We choose $\sigma^2$ based on the results from recent 3D MHD simulations of magnetic reconnection. For example,~\citet{Huang2016Turbulent} (Fig. 7 in their paper) showed that $\sigma^2$ could be 0.1--1 in a simulation with a guide field $B_g/B_0=1$. We expect the turbulence amplitude will be even larger when the guide field is weaker ($B_g/B_0=0.3$ in our simulations). Since the turbulence is driven by magnetic reconnection, we expect $L_c$ to be a fraction of the reconnection layer width and magnetic island sizes, which are both a few Mm in our simulations. A smaller $L_c$ means particles will experience more scattering and stay longer in the regions with strong flow compression. Due to the limitation of our MHD simulations, which have a relatively high plasma $\beta$ (about 0.1) and low Lundquist number ($10^5$ and $10^6$) compared to solar flares ($\sim 10^{12}$), we choose the lower end of $L_c\approx140$ km to achieve more efficient acceleration. Lower plasma $\beta$ and higher Lundquist number will likely lead to stronger flow compression and particle acceleration. Normalized by $\kappa_0=L_0v_\text{A0}\approx6.5\times10^{14}\text{ m}^2\text{ s}^{-1}$, $\tilde{\kappa}_{\parallel0}=\kappa_{\parallel0}/\kappa_0=1/16384$ in the simulations. We set $\kappa_\perp/\kappa_\parallel=0.01$, as test-particle simulations have suggested that particle parallel diffusion is much faster than cross-field diffusion~\citep{Giacalone1999Transport}.

We collect the MHD fields every 0.01 Alfv\'en crossing time $\tau_A$ and use them to solve the transport equation. Pseudo-particle electrons are evolved according to a set of stochastic differential equations (SDEs) equivalent to the Fokker-Planck form of the Parker transport equation~\citep{Zhang1999Markov,Florinski2009Four,Pei2010General,Guo2010,Li2018Large,Kong2019,Kong2020ApJ}. They are injected with an initial energy of 10 keV in regions with strong current density every $0.01\tau_A$. These injected particles have a speed about two times the thermal speed of the heated plasma in the flare region~\citep[$\sim 25$ MK,][]{Warren2018ApJ,Polito2018ApJ,Chen2021Energetic}. We have also tried different injection energies from 5 to 20 keV, and the results are similar. Note that these are not physical particles but represent a chunk of the distribution function $f$. We refer the interested readers to~\citet{Li2018Large} for implementation details of solving the SDEs.

\section{Simulation Results}
\label{sec:results}

\subsection{MHD Dynamics}

\begin{figure}[ht!]
  \centering
  \includegraphics[width=\linewidth]{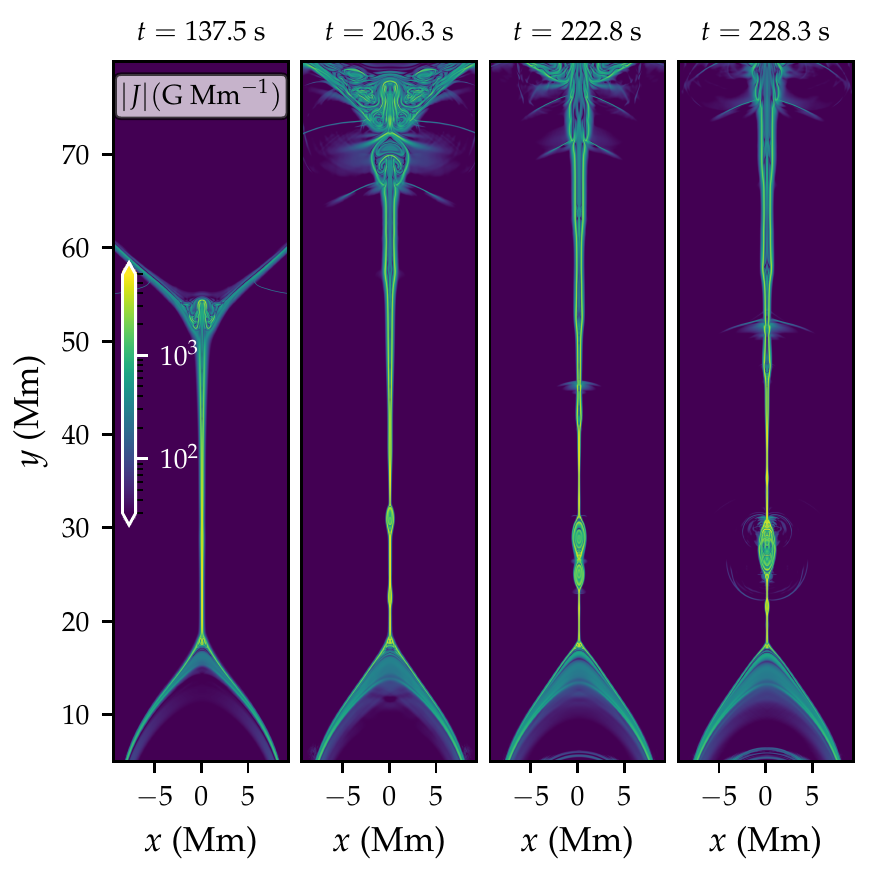}
  \caption{\label{fig:flare_s1e6}
  Time evolution of the current density in the run with $S=10^6$. Only the bottom and central region of the simulation domain is shown here. The time starts from the initialization of the simulation.
  }
\end{figure}

The large-scale evolution (e.g., formation of current sheet and eruption of the flux rope) is similar in both MHD runs. As the simulation starts, the rise of the flux rope stretches the field lines beneath and leads to the formation of an elongated current sheet. The reconnection process in the current sheet allows the flux rope to erupt from the solar surface (see Appendix~\ref{app:flare_evolution} for the global evolution of the flare region in the run with $S=10^5$). The two MHD runs are different in the structure of the reconnection region. In the run with $S=10^5$, the reconnection layer is Petschek-like~\citep{Petschek64}, including one reconnection X-point and an elongated reconnection exhaust, which is bounded by slow-mode shocks. The current density peaks in the current sheet and along the exhaust boundaries (Fig.~\ref{fig:flare_evolution}). When we increase the Lundquist number (or lower the resistivity $\eta$), the current sheet thins down to a smaller scale, resulting in a stronger current density. The pressure increase due to the resistive heating ($\eta J^2$) can then balance the inflow magnetic pressure. Consequently, the tearing instability quickly grow and lead to the formation of a series of magnetic islands (or plasmoids)~\citep{Shibata2001Plasmoid,Loureiro2007Instability,Bhattacharjee2009Fast}. Fig.~\ref{fig:flare_s1e6} shows the eruption region in the run with $S=10^6$, where the exhaust region collapses to a thinner and much more elongated current sheet early in the simulation (left panel). Later in the simulations, the current sheet breaks into multiple islands (right panels). Two of the islands ($y\sim 25$ Mm, each growing to a few Mm size) merge to form a larger island in a few seconds. We expect the merging region will efficiently accelerate energetic particles (see discussion below). Some even larger magnetic islands are ejected upward and collide with the erupted flux rope ($y > 70$ Mm). Plasma in these regions could be compressed (sometimes a shock may form; see~\citet{Kong2022Model}), and particles can be accelerated due to compression.

\begin{figure*}[ht!]
  \centering
  \includegraphics[width=\linewidth]{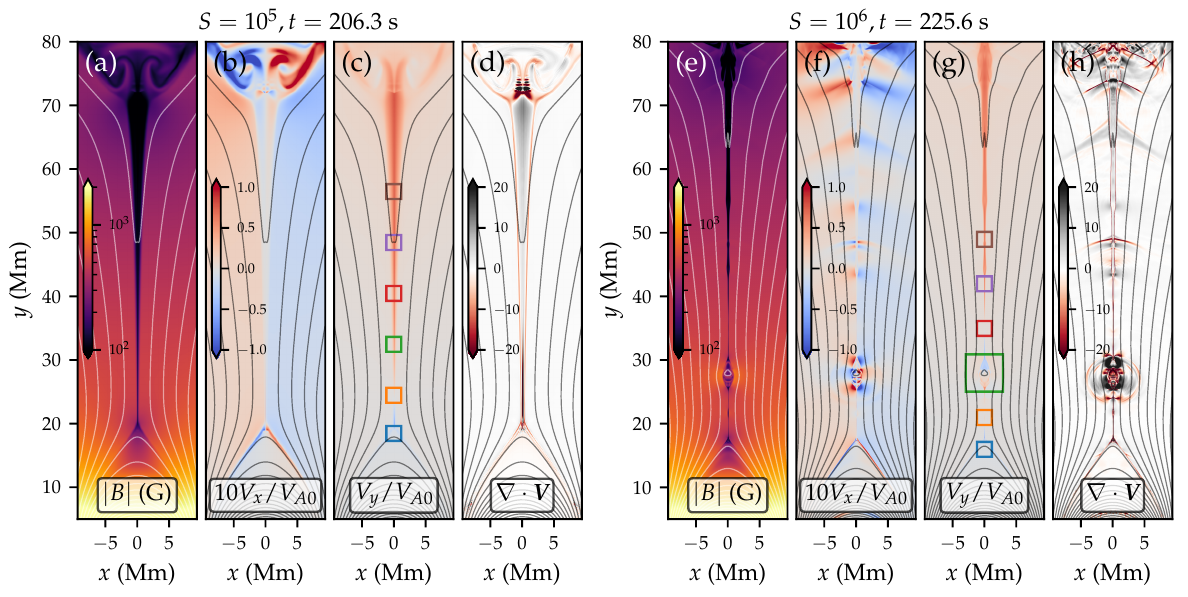}
  \caption{\label{fig:vel_mag}
  Snapshots of the reconnection layer in the two MHD runs. (a) \& (e) Magnetic field magnitude in Gauss. (b) \& (f) The $x$-component of the flow velocity. The plotted is $10V_x$ normalized by the Alfv\'en speed $V_{A0}$. (c) \& (g) The $y$-component of the plasma flow velocity. (d) \& (h) Flow compression $\nabla\cdot\vect{V}$. The boxes in (c) \& (g) indicate regions to obtain the local energy spectra shown in Fig.~\ref{fig:local_spectra}. The blue boxes are at the above-the-loop-top regions. The orange boxes cover the primary X-point. The red box in (g) covers another X-point; the large green box in (g) covers the island-merging region; the other boxes are distributed along the reconnection exhaust.
  }
\end{figure*}

Fig.~\ref{fig:vel_mag} shows the snapshots of the reconnection layer, including the magnetic field, inplane velocity field, and flow compression. The magnetic field is a few hundred Gauss outside the reconnection layer and about 100 Gauss (including the guide-field component) in the reconnection layer due to magnetic dissipation (panels (a) \& (e)). The magnetic field strength is consistent with that derived from the microwave observations by EOVSA~\citep{Chen2020Measurement}. The vertical cut through the center of the reconnection layer shows that the magnetic field is finite (not shown here), which is a result of the moderate guide field we include in the initial model setup ($\sim0.3B_0$;~\citet{Chen2020Measurement}) in the formed current sheet. In the run with $S=10^6$, the magnetic field can be compressed to several hundred Gauss in islands or island-merging regions (Fig.~\ref{fig:vel_mag} (g)). Additionally, magnetic field is weak at the Y-point ($y\sim 20$ Mm) near the flare looptop region in both cases. Such field configuration forms a ``magnetic bottle'' structure that is favorable in trapping energetic electrons at the looptop region \citep{Chen2020Measurement}, leading to a stronger HXR or microwave emission or possibly facilitating electron acceleration by plasma turbulence or termination shocks~\citep{Petrosian2012Stochastic,Chen2015Particle,Kontar2017,Fleishman2020Sci,Kong2019,Kong2020ApJ}.

\begin{figure}[ht!]
  \centering
  \includegraphics[width=\linewidth]{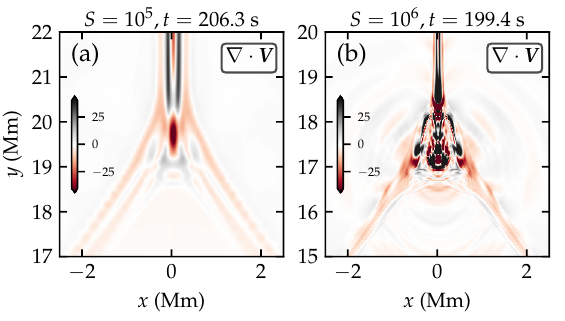}
  \caption{\label{fig:comp_looptop}
  Compression near the looptop region in the run with (a) $S=10^5$ and (b) $S=10^6$. Note that at $t=199.4s$, the current sheet starts to break into magnetic islands in the run with $S=10^6$.
  }
\end{figure}

The velocity plots reveal the large-scale reconnection inflow $V_x$ and outflow $V_y$ in the reconnection region. The inflow is about $0.01V_{A0}$, indicating that the reconnection rate is about 0.01. $V_x$ can reach $0.1V_{A0}$ in some regions (e.g., near the slow-mode shocks driven by Alfv\'enic magnetic island motions) in the run with $S=10^6$. The upward outflow can reach $V_{A0}$, but typically, the downward outflow is slower than $V_{A0}$. The reason is that the major X-point (in the orange boxes in (c) \& (g)) is close to the Y-point, resulting in a distance too short for the downward exhaust to open up and for the outflow to grow. The field structure and the slow downward outflow also explain the absence of a termination shock in the above-the-loop-top region in this simulation setup. When $S=10^5$, flow compression is only possible when either the reconnection inflow or outflow terminates since there is a single X-point (Fig.~\ref{fig:vel_mag} (d)). When the inflow terminates, it forms the slow shocks bounding the reconnection exhaust the compression region near the X-point. The outflow terminates either at the looptop region (see Fig.~\ref{fig:comp_looptop}(a) for a zoom-in view) or below the erupted large flux rope. These compression regions could accelerate particles when these particles are trapped by turbulence in these regions~\citep{Li2018Large,Kong2019,Kong2020ApJ}. In the run with $S=10^6$, the thin current sheet and the compressed magnetic field are accompanied by a stronger flow compression in the reconnection layer (Fig.~\ref{fig:vel_mag} (h)), especially when islands merge ($y\sim 28$ Mm). These regions will be efficient in accelerating particles (see discussion below). Additionally, the plasmoids also lead to slow shocks and magnetosonic waves propagating in the inflow region, which could help pre-heat the plasmas before entering the reconnection region~\citep{Arnold2021Slow}.

Fig.~\ref{fig:comp_looptop} shows the zoom-in views of the flow compression near the looptop region, where most nonthermal coronal emissions are observed~\citep{Masuda1994Loop,Chen2020Measurement}. When $S=10^5$, flow is compressed above the flare loops when the reconnection outflow terminates and at the flare loop boundaries. There is a weak compression in the whole flare loop region due to the contraction of the flare loops. Fig.~\ref{fig:comp_looptop}(b) shows a stronger compression when $S=10^6$. Additionally, the region is more ``turbulent'', with complex structures and multiple compression and expansion regions, which can help scatter and trap energetic particles~\citep{Kong2019,Kong2020ApJ}. We expect particles energized in the reconnection layer to accelerate further and thus boost the high-energy electron fluxes and emissions in the looptop region.

\begin{figure*}[ht!]
  \centering
  \includegraphics[width=0.75\linewidth]{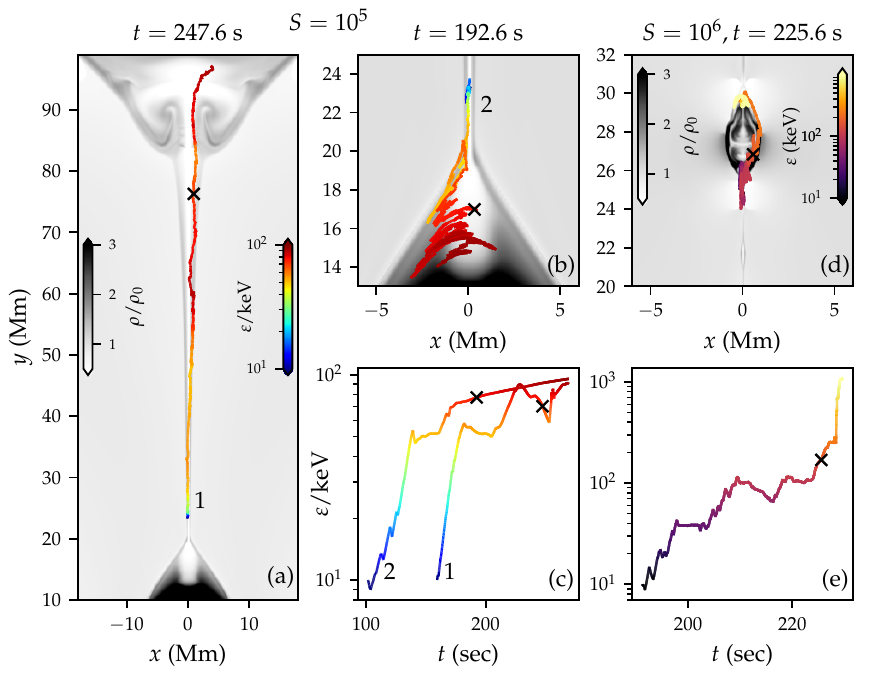}
  \caption{\label{fig:ptl_trajs}
  Pseudo particle trajectories. The left panels show the two trajectories in the run with $S=10^5$. The right panels show one trajectory in the run with $S=10^6$. (a) \& (b) Particle trajectories plotted on the plasma density contour. The trajectories are color-coded by their energies. The black crosses indicate the particle positions at the background time frames. (c) Time evolution of the particle kinetic energy. (d) The trajectory of one particle energized primarily near the island-merging region. The black cross indicates the particle position at $t=225.6$ s. (e) The corresponding time evolution of the particle kinetic energy.
  }
\end{figure*}

\subsection{Energetic Electrons in the Flare Region}

Particles gain energy in compression regions and are advected by the reconnection outflow up to the erupted flux rope and down to the flare loops. Although they are not ``real'' particles, the pseudo particles can still reveal where the acceleration occurs. Fig.~\ref{fig:ptl_trajs} shows some pseudo particle trajectories typical in the simulations. In general, particles can be accelerated in the current sheet, the loop-top region, and the islands-merging region. Particles 1 and 2 are injected near the X-point. The former gains energy near the X-point where reconnection inflow terminates and compression is large (Fig.~\ref{fig:vel_mag} (d)). Its energy increases from 10 keV to about 50 keV in about 20 s. The particle gets advected upward by the reconnection outflow and gains energy up to 100 keV until it reaches the erupted flux rope. The initial acceleration of particle 2 is similar to particle 1. The difference is that particle 2 is advected downward to the above-the-loop-top region, where the particle is trapped in the weak field region. It is slowly accelerated by the weak compression when the flare loops contract (Fig.~\ref{fig:comp_looptop}(a)). Fig.~\ref{fig:ptl_trajs}(d) shows one particle trajectory in the run with $S=10^6$. This particle is primarily energized near the island-merging region. It continuously gains energy up to MeV when it diffuses across the island-merging region, where the flow compression is strong. This result confirms our earlier anticipation that the island-merging regions are efficient in accelerating particles~\citep{Drake2013Power,Arnold2021PRL}.

\begin{figure*}[ht!]
  \centering
  \includegraphics[width=\linewidth]{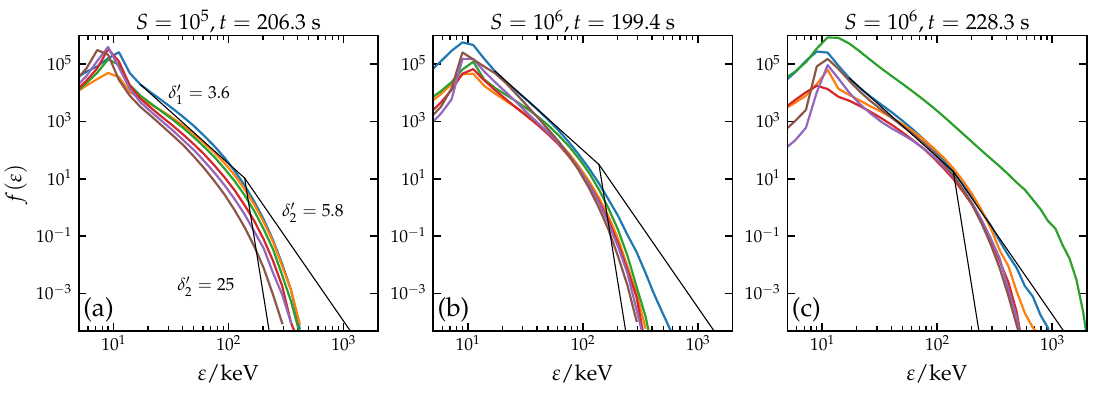}
  \caption{\label{fig:local_spectra}
  Local electron energy spectra along the reconnection layer. (a) The spectra in the colored boxes shown in Fig.~\ref{fig:vel_mag}(c). The black lines indicate the broken power-law spectra derived from the microwave emissions in the above-the-loop-top source during the early impulsive phase of the Sep. 10th, 2017 flare~\citep{Chen2021Energetic}. $\delta_1^\prime$ is the lower power-law index, and $\delta_2^\prime$ is the higher power-law index. The two values of $\delta_2^\prime$ indicate the hardening of the higher power-law in 20 s. (b) The spectra in the run with $S=10^6$ at $t=199.4$ s, when the current sheet just starts to break into islands. The boxes are distributed similarly as Fig.~\ref{fig:vel_mag}(c). (c) The spectra in the colored boxes shown in Fig.~\ref{fig:vel_mag}(g). We choose the time frame $t=228.3$, which is 2.7 s later than Fig.~\ref{fig:vel_mag} (e)--(h), to show the effect of compression acceleration in the island-merging region (green).
  }
\end{figure*}

Fig.~\ref{fig:local_spectra} compares the local electron energy spectra in the transport modeling using the two MHD runs. Fig.~\ref{fig:local_spectra}(a) shows the spectra in the run with $S=10^5$. The electrons are accelerated to a few hundred keV, and their distributions break at about 100 keV. Although the spectral shapes are similar, they have some differences. The spectrum in the above-the-loop-top region (blue) has the highest flux, and its low-energy part is close to the observed spectrum derived from microwave emissions~\citep{Chen2021Energetic}. However, the high-energy distribution ($> 100$ keV) does not show a hardening similar to the observations, suggesting that the compression in the looptop region in our simulation is inefficient in further accelerating particles. Therefore, additional acceleration mechanisms (termination shocks or turbulence) in the looptop region may be necessary to boost the high-energy fluxes~\citep{Chen2015Particle,Kong2019,Kong2020ApJ,Fleishman2020Sci}. The spectra near the X-point (green and orange) are harder (lower low-energy flux but similar high-energy flux) than those in the reconnection exhausts, consistent with those derived from spatially resolved microwave spectral analysis~\citep{narukage2014,Chen2020Measurement}. Fig.~\ref{fig:local_spectra}(b) \& (c) show similar results in the run with $S=10^6$ at two time frames. Overall, electrons are accelerated to higher energies because of the stronger flow compression than $S=10^5$ (Fig.~\ref{fig:vel_mag}(d) \& (h)). Electrons in all regions can reach a few hundred keV and up to several MeV in some regions (Fig.~\ref{fig:comp_looptop}(c)). At $t=199.4$ s (Fig.~\ref{fig:local_spectra}(b)), the spectra are similar to those in the run with $S=10^5$ except in the looptop region (blue line) before magnetic islands are produced and dominate particle acceleration. The spectrum in the looptop region has the highest fluxes, especially for electrons $> 300$ keV, consistent with the observations showing that the high-energy electron fluxes peak in the looptop region and sharply decrease toward the X-point~\citep{Chen2020Measurement}. The enhancement of the high-energy electron fluxes suggests additional acceleration due to compression is more efficient in this run than when $S=10^5$, as expected from Fig.~\ref{fig:comp_looptop}(b). After the current sheet breaks into magnetic islands (Fig.~\ref{fig:local_spectra}(c)), the spectra become harder in all regions due to the acceleration associated with magnetic islands, and the local spectra differ in different regions due to the complex structures and associated compression (Fig.~\ref{fig:vel_mag}(h)). Fig.~\ref{fig:vel_mag}(g) shows that the orange and red boxes are near the flow diverging regions and thus the X-lines. Particles in the two boxes have similar high-energy fluxes but lower low-energy fluxes than the others, resembling harder spectra for electrons $<100$ keV. In the looptop region (blue box), both low-energy and high-energy spectra resemble power-law distributions with power-law indices similar to the observed values~\citep{Chen2021Energetic}. The most significant difference between this run and $S=10^5$ is in the island-merging regions (green box). Fig.~\ref{fig:local_spectra}(b) shows that electrons can be accelerated over MeV and develop a harder ($p\approx 3$) and more extended power-law spectrum due to the stronger flow compression associated with island mergers. Depending on island sizes, the mergers can enhance high-energy electron fluxes in a few seconds or shorter, which is consistent with the short pulses observed in HXR and microwave emissions~\citep{Knuth2020}.

\begin{figure*}[ht!]
  \centering
  \includegraphics[width=0.7\linewidth]{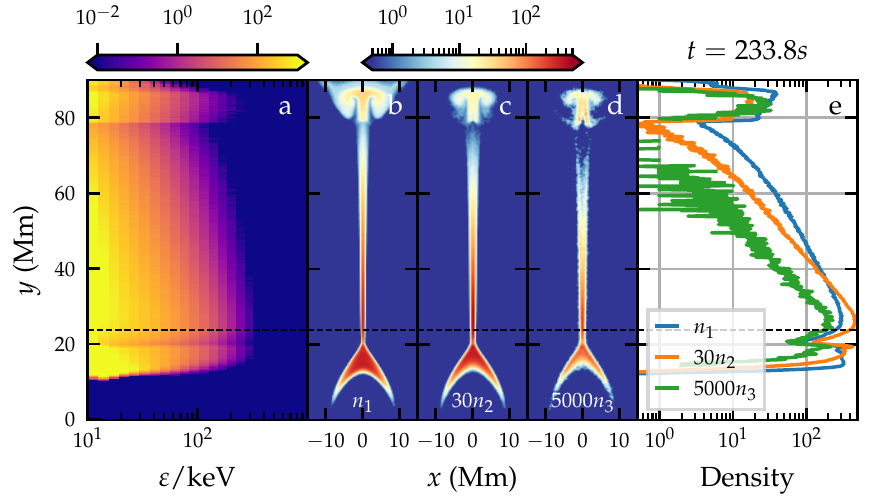}
  \caption{\label{fig:dist_maps_s1e5}
  Electron distribution maps at $t=233.8$ s in the run with $S=10^5$. (a) The stacked spectrum in the center of the reconnection layer. (b)--(d) Particle distributions in three energy bands: band 1 is from 30 keV to 100 keV; band 2 is from 100 keV to 300 keV; band 3 is over 300 keV. The higher energy bands are multiplied by different factors for better visualization. (e) The vertical cuts of the three bands through the center of the reconnection layer ($x=0$). The black dashed line indicates the height of the reconnection X-point.
  }
\end{figure*}

Since the particle transport model can be solved efficiently, we can use a large number of pseudo particles to get the energetic particle maps. Fig.~\ref{fig:dist_maps_s1e5} shows the electron maps for $S=10^5$. The stacked spectra (panels (a)) show that the electron flux peaks in the looptop region, the reconnection X-line, and the region below the flux rope, where the plasma is compressed (Fig.~\ref{fig:vel_mag}(d)). We divide particles into different energy bands and get the distribution maps shown in Fig.~\ref{fig:dist_maps_s1e5}(b)--(d). High-energy electrons can fill the flare reconnection region, consistent with the observations showing that the microwave emission generated by energetic electrons can fill the flare reconnection region~\citep{Chen2020Measurement}. Compared with $n_1$ (30--100 keV) and $n_2$ (100--300 keV), $n_3$ ($>$300 keV) is more concentrated at the lower part of the reconnection layer, consistent with the observations~\citep{Chen2020Measurement}. The vertical cuts (Fig.~\ref{fig:dist_maps_s1e5}(e)) show that the electron fluxes peak in the looptop region and the X-line and gradually decrease along the exhaust, which is inconsistent with observations showing that the electron flux peaks at the above-the-loop-top region and sharply decreases towards the reconnection X-point ((Fig. 3(d) in~\citet{Chen2020Measurement})). This result suggests that additional acceleration mechanisms, such as turbulence acceleration and termination shocks~\citep{Chen2015Particle,Kontar2017,Fleishman2020Sci,Kong2019,Kong2020ApJ}, might be necessary to boost the high-energy electron fluxes in the above-the-loop-top region to explain the observed microwave and HXR emissions.

\begin{figure*}[ht!]
  \centering
  \includegraphics[width=0.7\linewidth]{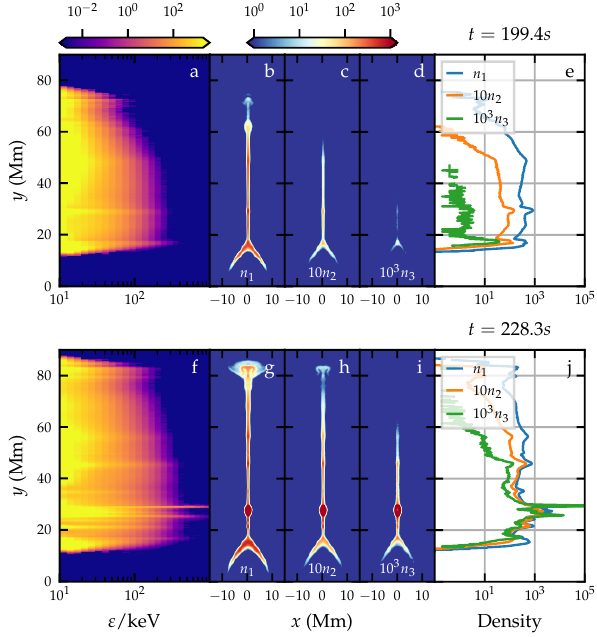}
  \caption{\label{fig:dist_maps_s1e6}
  Electron distribution maps similar to Fig.~\ref{fig:dist_maps_s1e5} in the run with $S=10^6$ at $t=199.4$ s (top panels) and $t=228.3$ s (bottom panels). $n_1$ is from 30 keV to 100 keV; $n_2$ is from 100 keV to 300 keV; $n_3$ is over 300 keV.
  }
\end{figure*}

Fig.~\ref{fig:dist_maps_s1e6} show similar results for the run with a higher Lundquist number $S=10^6$ at two time frames shown in Fig.~\ref{fig:local_spectra}(b) \& (c). The stacked spectra show that the electron fluxes develop more structures along the current sheet once magnetic islands are generated, consistent with the observed high-energy electron spectra derived from the microwave emissions (Fig. 3(d) in~\citet{Chen2020Measurement}). Comparing panels (a) and (f), we find that particles are continuously accelerated to higher energies, and they are advected and diffuse to broader regions. The electron maps (panels (b)--(d)) show that high-energy electrons can fill a significant portion of the reconnection region and higher energy bands are more concentrated in the lower region ($y<60$ Mm). Most dramatically, electrons with energy $> 300$ keV ($n_e$) are primarily in the looptop region, as expected from the local spectra (Fig.~\ref{fig:local_spectra}(b)). The vertical cuts (Fig.~\ref{fig:dist_maps_s1e6}(e)) show that $n_3$ in the looptop region is nearly two orders higher than in the reconnection region, demonstrating the efficient acceleration due to flow compression in the looptop region again. These results are consistent with the observations of the early impulsive phase of Sep. 10th 2017 flare~\citep{Chen2020Measurement}. Once magnetic islands grow substantially in size and play essential roles in particle acceleration, the model results differ from the observations. The stacked plot and electron distributions and its vertical cuts all show that high-energy electron fluxes  peak in the island-merging region (20 Mm $<y<$ 30 Mm) due to the strong compression acceleration. The fluxes in the other regions are much lower, which was not observed in the early impulsive phase of Sep. 10th flare. To explain the observations, additional acceleration mechanisms are required to accelerate the electrons further in the looptop region. The merger-associated acceleration could explain the more efficient particle acceleration in the main impulsive phase.


\section{Discussions and Conclusions}
\label{sec:con}
In this paper, we have studied electron acceleration and transport in the early impulsive phase of Sep. 10th 2017 flare by solving Parker's transport equation in the large-scale eruption region modeled by MHD simulations. We found that electrons can be accelerated to hundreds of keV due to plasma flow compression and develop power-law spectra similar to the observations~\citep{Chen2021Energetic}. The energetic electrons can fill a significant portion of the flare reconnection region and thus could explain the volume-filling microwave emissions~\citep{Chen2020Measurement}. Reconnection X-point, exhaust, and flow termination regions are possible sites for electron acceleration by flow compression. We found that the acceleration becomes stronger when magnetic islands dominate high-energy particle acceleration in MHD simulations with a higher Lundquist number $S=10^6$. The electrons can be accelerated to over MeV in contracting and merging magnetic islands due to strong flow compression in these regions. Comparing the two MHD runs with different Lundquist numbers, we found that the acceleration is stronger in the looptop region when $S=10^6$ than when $S=10^5$ because of the stronger compression and more complex field structures. The high-energy fluxes strongly peak in the looptop region before islands play the dominant roles in accelerating electrons when $S=10^6$, consistent with the observations in the early impulsive phase of this flare~\citep{Chen2020Measurement}.

In the transport model, electrons are accelerated to high energies through multiple energizations near reconnection X-lines, in reconnection exhausts, contracting and merging magnetic islands, and the looptop regions. Moreover, because of the scattering (spatial diffusion), particles can sample the same compression region multiple times and continuously gain energy, similar to the diffusive shock acceleration, where particles cross the shocks multiple times. The acceleration process is similar to~\citet{Guidoni2016ApJ,Guidoni2022ApJ} where the multi-phase acceleration is achieved through transferring particles from one plasmoid to another. They found that the model can reproduce the observed spectral parameters.

Energetic particle modelings based on both MHD simulations show some inconsistency with observations. In the MHD run with $S=10^5$, the high-energy electron fluxes peak in both the looptop region and the X-line region, indicating that the acceleration in the looptop region is not efficient enough to account for the observed high concentration of nonthermal electrons at the looptop. To explain the observations, we need additional acceleration mechanisms to further accelerate the electrons and boost the high-energy electron fluxes in the looptop region. Possible mechanisms include stochastic acceleration \citep{Petrosian2012Stochastic,Fleishman2020Sci}, termination shocks~\citep{Guo2012Particle,Chen2015Particle,Kong2019,Kong2020ApJ}, collapsing traps~\citep{Somov1997,karlicky2004}, and coalescing magnetic islands \citep{Drake2013Power}. Although electron acceleration is more efficient in the loop region in the run with $S=10^6$, the high-energy electron fluxes strongly peak in the island-merging regions after magnetic islands grow significantly in size and merge with each other. We expect the nonthermal emissions will be concentrated in the island-merging regions, but observations in the early impulsive phase of Sep. 10th flare do not show evidence of magnetic islands of a sufficient size nor strong localized nonthermal emissions in the current sheet. These observations are consistent with earlier 2.5D MHD simulations of eruptive flares with more realistic geometries and boundary conditions~\citep[e.g.,][]{Karpen2012ApJ,Guidoni2016ApJ}, which do not exhibit plasmoid mergers. However, we emphasize that the simulations usually have a Lundquist number (up to $10^7$) much smaller than in solar flares ($\sim 10^{12}$). Previous MHD simulations have suggested that the number of plasmoids increases with the Lundquist number~\citep[e.g.,][]{Samtaney2009,Huang2010Scaling,Loureiro2012}, and smaller plasmoids can be generated as the Lundquist number increases. Although large plasmoid mergers are not often observed in solar flares, merging between smaller plasmoids (below the observation resolutions) might occur. As our model does not resolve these small-scale structure, we do not know whether the small mergers or the large mergers are more important in electron acceleration, which requires more studies in the future. It is worth mentioning that both $S=10^5$ and $10^6$ runs in this work may significantly underestimate the plasma compression in the flare loop top region. Because of the high density (and plasma $\beta$) setting and limitation of computational resources, we can not reveal the clear termination shock and highly turbulent post-shock plasma features as shown in other similar simulations~\citep{Kong2019,Shen2018,Shen2022}.

In this paper, we focus on the early impulsive phase of the 2017 Sep. 10th flare, when the nonthermal fraction of energetic electrons~\citep[$\sim$1\%,][]{Chen2021Energetic} is much lower than that in the main impulsive phases of some of the largest flares~\citep{Krucker2010Measure,Krucker2014Particle}. Therefore, the ``number problem''~\citep[e.g.,][]{Brown1977,Fletcher2007} is not critical for the current study. As discussed in the simulation setup, we inject particles in regions with large current densities, including both the X-points and reconnection exhausts. Thus, only part of the injected particles crosses the reconnection X-points. Most of them enter the reconnection region through the reconnection exhaust boundaries, similar to the Petschek model. Fig.~\ref{fig:vel_mag}(d) shows that the flow compression is weak in most regions in the reconnection exhaust. As a result, the electrons injected near the reconnection X-point ($y=$ 20--30 Mm) are more efficiently accelerated than those injected through the exhaust boundaries. Consequently, to the end of the simulation, only a few percent of the electrons are accelerated to over tens of keV when there is only one X-point (low Lundquist number case). When the Lundquist number is higher, the current sheet can break into multiple plasmoids, and many more particles can be accelerated by plasmoid compression, which could help resolve the ``number problem'' in the main impulsive phase.

The assumptions we made in solving the transport model might need justifications and further support from the observations and simulations. First, the Parker transport equation assumes a nearly isotropic particle distribution, which could be reasonable if reconnection-driven turbulence~\citep{Daughton2011Role,Guo2015Efficient,Dahlin2015Electron,Li2019Formation,Zhang2021Efficient} or instabilities~\citep{Roberg-Clark2019ApJ} can strongly scatter the energetic particles. When the particle distribution is nearly isotropic, the first-order acceleration term is due to flow compression, which has been demonstrated to be equivalent to the acceleration associated with particle guiding-center drift motions~\citep{Dahlin2014Mechanisms,LeRoux2015Kinetic,Li2017Particle,Li2018Roles}. However, suppose the scattering is not frequent enough. In that case, particle distributions will be anisotropic, and we need to solve models like the focused transport equations~\citep{Zank2014Particle,LeRoux2015Kinetic} or the drift kinetic equation~\citep{Drake2013Power,Montag2017Impact} instead and include additional acceleration mechanisms (e.g., flow shear,~\citet{Earl1988Cosmic,LeRoux2015Kinetic,Webb2018,Webb2019}).

Second, the transport modeling does not include feedback from energy electrons to the background plasma flow. This assumption is reasonable for the early impulsive phase of Sep. 10th flare because the observations suggest that the nonthermal particle fraction is small~\citep[$\sim$1\%,][]{Chen2021Energetic}, and therefore the feedback may not be critical. However, some observations suggest that up to 50\% of the released magnetic energy could be in the nonthermal electrons~\citep{Lin1976Nonthermal,Aschwanden2017}, which could modify the reconnection dynamics and change the particle acceleration processes in flares. In that case, the feedback of energetic particles through energy (or pressure) exchange or current density should be included in the model~\citep{Bai2015ApJ,Drake2019,Arnold2019,Arnold2021PRL}.

Third, although the observations indicate plasma is turbulent in the flare region, the turbulence properties (e.g., turbulence amplitude, spectrum, correlation length, anisotropy) relevant to particle acceleration are unknown. The reconnection-driven turbulence is under active research using 3D MHD~\citep{Huang2016Turbulent,Kowal2017Statistics} and kinetic simulations~\citep{Daughton2011Role,Li2019Formation,Guo20213D}, and we expect to see more understanding of these turbulence properties. This progress can lead to more accurate spatial or energy diffusion coefficients than estimates using the quasi-linear theory~\citep{Jokipii1971Propagation,Giacalone1999Transport}. Besides, the turbulence properties could be spatially dependent. The current sheet, ambient coronal plasma, looptop, and flare loops might all have different turbulence properties. We expect the turbulence to be stronger in the reconnection layer due to the reconnection activities. Modeling the transition of the turbulence properties from the ambient coronal plasma to the reconnection layer requires a future study.

Fourth, more physics should be included to compare with the solar flare observations, such as thermalization and collisional loss~\citep{Fletcher1995AA,Holman2011,Battaglia2012,Jeffrey2014PhDT}. These are particularly important in high-density regions (e.g., looptop regions and flare loops), where frequent collisions will lead to stronger collisional loss and affect electron transport. The electron distributions will be more isotropic due to the collisions, which will help trap energetic electrons in the looptop region to form the coronal HXR microwave emissions and modify the precipitation of energetic electrons to the footpoints.

Fifth, the plasma parameters (e.g., plasma $\beta$ and guide field) that can affect particle acceleration and transport should be modeled more accurately in MHD simulations for studying particle acceleration in solar flares. According to kinetic simulations~\citep{Li2018Roles}, flow compression is stronger in the lower-$\beta$ and lower guide-field reconnection region. We expect the acceleration due to flow compression will be more efficient in these reconnection layers. Last but not least, it is important to realize that coronal reconnection and nonthermal emission can involve significant 3D effects. The observations are often the projection of the 3D signatures onto a plane~\citep{Chen2020ApJ}. To explain the observations and further understand electron acceleration, it is requisite to build a 3D model of nonthermal electron acceleration and associated emissions during solar flares.

To conclude, by solving the Parker transport equation in the large-scale flare region modeled by MHD simulations, we model the electron acceleration in the early impulsive phase of Sep. 10th 2017 flare. We found that electrons can be accelerated up to MeV due to flow compression in various regions and fill a significant portion of the reconnection region, both of which are consistent with observations. The model can reproduce the spectral shape and high-energy electron maps before magnetic islands dominate high-energy electron acceleration. These electron maps can be used for modeling nonthermal emissions directly comparable to the observations.

\acknowledgments
We thank the anonymous referee for very helpful and constructive review. X.L. acknowledges the support from NASA through Grant 80NSSC21K1313, National Science Foundation Grant No. AST-2107745, and Los Alamos National Laboratory through subcontract No. 622828. F.G. acknowledges the support from NASA grants 80HQTR20T0073, 80HQTR20T0040, 80HQTR21T0087, 80HQTR21T0103, NSF grant AST-2109154, and the support from LDRD program at LANL. B.C. acknowledges the support from NASA grant 80NSSC20K1318 and NSF grants AGS-1654382, AST-2108853 to NJIT. C.S. acknowledges the support from NASA through Grant 80NSSC21K2044, National Science Foundation Grant No. AST-2108438. Simulations were performed at the National Energy Research Scientific Computing Center (NERSC) and the Texas Advanced Computing Center (TACC) at The University of Texas at Austin. This work was facilitated in part by the NASA Drive Science Center on Solar Flare Energy Release (SolFER).

\appendix
\section{Flare evolution}
\begin{figure}[ht!]
  \centering
  \includegraphics[width=\textwidth]{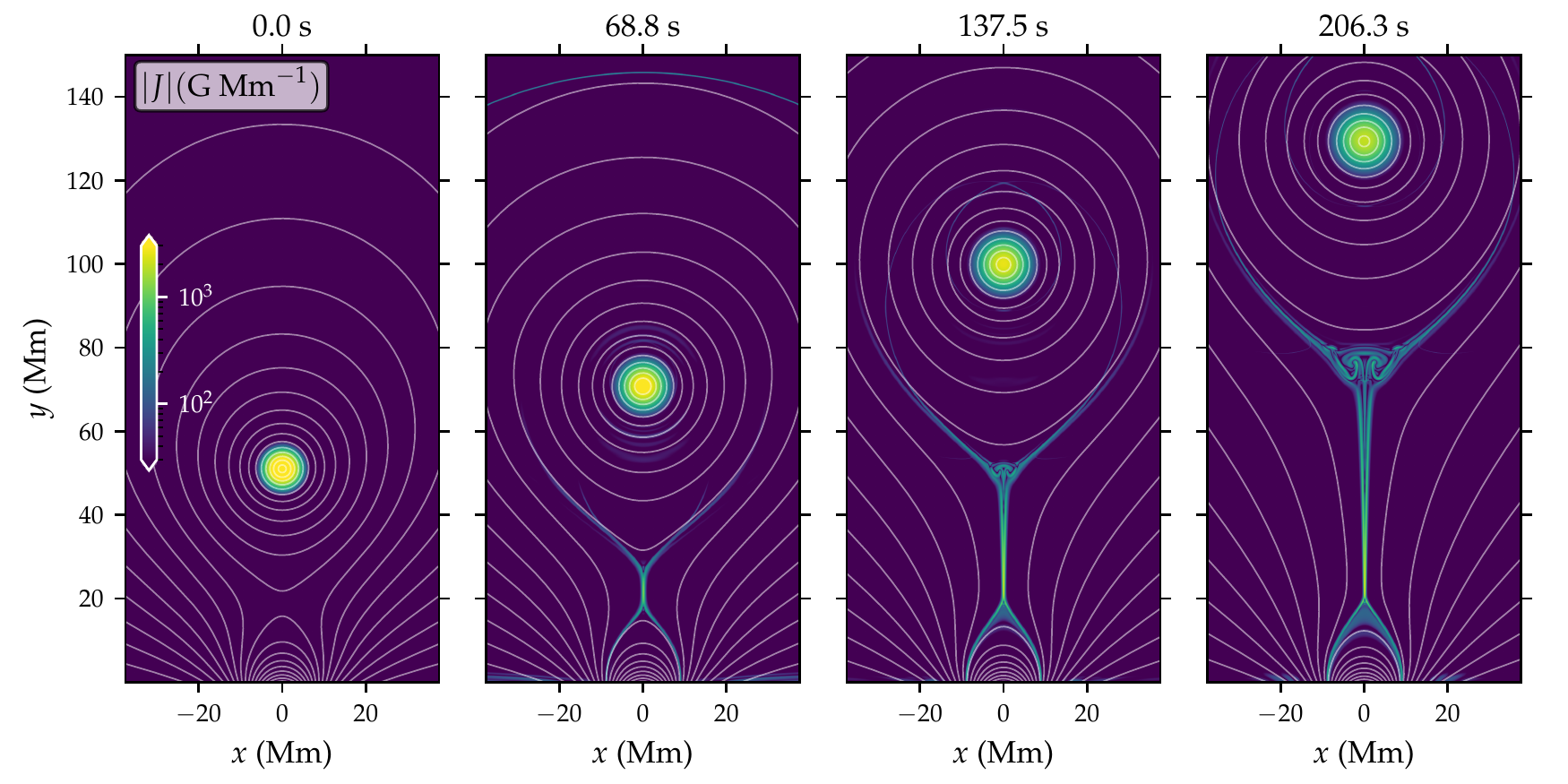}
  \caption{\label{fig:flare_evolution}
  Time evolution of the current density in the flare region when $S=10^5$.
  }
\end{figure}

Fig.~\ref{fig:flare_evolution} shows the time evolution of the flare region in the MHD simulation with $S=10^5$. As the flux rope rises, the magnetic field lines become stretched because the magnetic field lines are line-tied to the photosphere. An elongated current sheet develops below the flux rope, and the current density inside continues to intensify. The magnetic field lines start to reconnect, and the magnetic fluxes are continuously brought into the current sheet by the reconnection inflow. This process continually cuts magnetic field lines connecting to the flux rope, allowing it to erupt (see more details in~\citet{Chen2020Measurement}).
\label{app:flare_evolution}

\bibliography{references}{}
\bibliographystyle{aasjournal}
\end{document}